\documentclass[10pt]{article}

\usepackage{amsmath}
\usepackage{amssymb}

\usepackage{graphicx}

\usepackage{cite}

\usepackage{color} 

\usepackage{setspace} 
\usepackage[labelfont=bf,labelsep=period,justification=raggedright]{caption}

\makeatletter
\renewcommand{\@biblabel}[1]{\quad#1.}
\makeatother

\date{}

\graphicspath{{../Figures/}}

\begin{document}

\begin{flushleft}
{\Large
\textbf{Plasticity facilitates sustainable growth in the commons}
}

Matteo Cavaliere$^1$ and Juan F. Poyatos$^{1*}$\\

\bf $^1$ Logic of Genomic Systems Laboratory (CNB-CSIC), Madrid, Spain
\end{flushleft}

\section*{Abstract}


In the commons, communities whose growth depends on public goods, individuals often rely on surprisingly simple strategies, or heuristics, to decide whether to contribute to the common good (at risk of exploitation by free-riders). Although this appears a limitation, here we show how four heuristics lead to sustainable growth by exploiting specific environmental constraints.  The two simplest ones --contribute permanently or switch stochastically between contributing or not-- are first shown to bring sustainability when the public good efficiently promotes growth. If efficiency declines and the commons is structured in small groups, the most effective strategy resides in contributing only when a majority of individuals are also contributors. In contrast, when group size becomes large, the most effective behavior follows a minimal-effort rule: contribute only when it is strictly necessary. Both plastic strategies are observed in natural systems what presents them as fundamental social motifs to successfully manage sustainability.

\newpage
\section*{Introduction}
In many biological, social and economic systems, there exists a continuous interplay between individual actions and collective dynamics~\cite{Levin1999,Conradt2005,Kagel1995}. This interplay becomes particularly significant when the individual decisions on how to contribute to a public resource determine in the end the sustainability of the whole. The choice of contributing --that implies personal costs-- favors community growth but also promotes the appearance of {\em free-riders}.  These agents take advantage of the public good (PG), spread in the population and eventually bring its collapse~\cite{Hardin1968}.

This predicted scenario does not correspond however with many observations. Stable communities whose growth is based on PGs are indeed observed at all scales; from microbial aggregates, e.g., biofilms depending on the individual contribution of extracellular substances, e.g.,~\cite{Diggle2007,Nadell2008},  to human commons, e.g., fisheries, forests, etc (note that in some of these cases the choice is not so much  to contribute but to make appropriate use of a shared resource)~\cite{Ostrom2005,Bowles2011}.  These findings triggered the interest on understanding the type of behavioral strategies that could be adopted by individuals to help avoid collapse and how such outcome could further depend on specific structural features characterizing the community.

Notably, the adoption of simple strategies appears to be efficient enough to promote sustainability~\cite{Nadell2008,Rustagi2010}. Simple rules contrast the idea of elaborated behaviors that allow individuals to optimally maximize benefits, a null model particularly extended in studies of human commons. In this context, the relevance of elementary strategies (or heuristics) was first investigated by Herbert Simon that also pioneered the essential connection between heuristics and the particular environment where they are to be applied~\cite{Simon1957}  (see also~\cite{Ostrom1994, Gigerenzer2002,Gigerenzer2011}). In a broader perspective, heuristics -- sometimes interpreted as behavioral ``limitations''-- can then represent effective strategies to deal with complex ecological constraints --a consideration that applies to bacterial, animal and human decision-making circumstances~\cite{Nadell2008,Marsh2002,Agrawal2001,Ostrom2005,Rustagi2010}.

In this manuscript, we examine how simple strategies, that exhibit limited information processing, can nevertheless be adjusted to exploit certain environmental constraints to attain sustainable growth. We model the commons by means of a stylized ecological public-good model in which a population of agents is organized in groups (where they are involved in a public-good game) and the supply of PG determines population density ~\cite{Hauert2006} (Fig.~S1). Core structural factors characterizing the ``community'' can also be modified in the model, such as group size $N$ or resource characteristics, i.e., PG efficiency, $r$.

We additionally consider three types of individual strategies of increasing sophistication (Fig.~1). First, we examine the validity of the simplest one, i.e., permanent production of PG. We find that this strategy works when PG creation efficiently induces growth and the commons is structured in relatively small groups. Then, we assume a strategy that stochastically alternates between contribution and non-contribution. This rule, as simple as it is, enlarges the range of commons where its adoption leads to sustainability (as compared to the previous case). Finally, we introduce a heuristic in which a simple sensing mechanism is at work. This sensing allows individuals to condition their contribution on the composition of their last interaction group, i.e., it involves some degree of behavioral plasticity. We discover two opposite plastic heuristics to be efficient in two contrasting environmental situations. While {\em positive} plasticity (contribute only when most individuals in the group are contributing) works for low efficiency and small groups, {\em negative} plasticity (contribute merely when it is strictly necessary) does it for high efficiency and large groups.

\section*{Results }
\subsection*{Constitutive Production and the Risks of the Commons} 
We first examined the consequences of the simplest possible strategy: permanent and indiscriminate production of PG. This strategy maximizes growth but additionally favors the emergence of cheaters (that arise by mutation from an all$P$ population). Cheaters rapidly invade a resident population but also cause its decline, because of the coupled decay in PG (less $P$s) that limits growth (Fig.~2$A$). 

The population cascade associated to cheater expansion can unexpectedly direct to its recovery. This is linked to the group structure of the interactions in the commons. Sufficiently small density causes the appearance of groups primarily composed of only $P$s or only cheaters (Figs.~2$A$-$B$) that multiplies the replication of constitutive $P$ and reduces that of cheaters, both processes contributing to the recovery of the population (when enough inter-group composition variance is generated, in what is known as the Simpson's paradox~\cite{Sober1984,Hauert2006}, see also Supplement). 

This recovery dynamics includes however an added risk, since the low density could precipitate population extinction by stochastic demographic effects~\cite{Melbourne2008}, Figs.~2$A$ and 2$C$. Risk is raised when the population repeatedly exhibits {\em critical cascades}, i.e., declines in density below a particular minimal value. The final outcome between recovery and extinction is strongly determined by the intrinsic properties of the commons. Constitutive production reveals in this way as a successful strategy when the PG efficiently determines growth ($r$ sufficiently high, Figs.~2$D$-$F$) or when groups within the commons are relatively small (controlling for $r$, Fig.~S2).

\subsection*{Stochastic Production Can Reduce the Risks} 
We analyzed a second strategy in which individuals choose randomly whether to contribute or not to the PG (i.e., they can sometimes decide to free ride).  Specifically, agents present a $nP$ state with probability $q_>$ (or, conversely, an $P$ state with 1-$q_>$, see Fig.~1). Therefore, stochastic producers are totally unable to sense the composition of their interaction group (the amount of available PG).

A homogeneous population of stochastic producers generates a constant sub-population of $nP$s that decreases PG levels with two consequences. It can reduce the chances of cheaters to replicate (which favors sustainability) but also drive the system to extinction even without any cheater present --by cause of a severe PG reduction. We quantified this trade-off by computing the number of critical cascades, as before, and their duration (i.e., number of consecutive steps below the minimal density threshold). Increasing $q_>$ reduces the number of critical cascades but also increases steadily cascade duration (that reflects the delay in the appearance of all$P$ groups associated to population recovery). 

This trade-off indicates an optimal $q_>$ that minimizes the frequency of extinctions and defines the exact stochastic rate for a successful strategy (Figs.~3.$A$-$B$). When agents follow this optimal stochastic heuristic, cheater replication is limited event for high densities (generally below 50$\%$, Figs.~3$C$-$D$), population oscillations are damped,  and extinction risks reduced. This dynamics contrasts with the constitutive heuristics scenario in which cheater replication can reach very high values in dense populations (Fig.~S3).

Sustainability is thus attained in a wider range of commons (in terms of $r$ and $N$) when agents followed a stochastic strategy. However, this range is still limited. A relative decrease in $r$ (Fig.~S4) or an increase in $N$ (Fig.~S5) implies once more an increment in the number of critical cascades and in this way of extinctions. Alternative strategies are required in those commons.

\subsection*{Plastic Cooperation Favors Sustainable Growth} 
To analyze if the addition of some basic information-processing features could direct to more effective heuristics (in commons where growth is unsustainable with the use of constitutive or stochastic production) we examined a third strategy that includes a simple sensing mechanism. This sensing permits agents evaluate the relative abundance of $P$s in the group where they most recently played the game (this can be estimated by means of the amount of PG received).

If the PG obtained in the previous interaction is above (below) a particular threshold $\theta$, individuals exhibit the $nP$ phenotype with probability $q_>$ ($q_<$). This defines a general plastic producer, see Fig.~1. We studied then the dynamics of a population of individuals exhibiting different plastic heuristics (distinct $q_>$, $q_<$, $\theta$) in a range of commons (characterized by $r$ and $N$). By scanning the diverse heuristic/commons conditions, we were able to identify two specific plastic heuristics that leads to sustainable growth in wider range of commons.

\subsubsection*{Positive Plasticity is the Most Effective Strategy for Small Groups and Low Efficiency} 
The most valuable heuristic in commons characterized by small groups and low PG efficiency $r$ consists on contributing to PG only if most members of the agent's recent interaction group were also contributing. Individuals that follow this heuristic (positive plastic producers) immediately react to the presence of $nP$ (or cheaters) in their past interaction, becoming $nP$ themselves (formally, they present a small $q_>$, but a large $q_<$ and $\theta$, see Fig.~1). The appearance of cheaters in a population of positive plastic producers (in state $P$) originates consequently an immediate decrease of PG in each group which triggers the remaining $P$s to stop contributing and switch to $nP$ (Fig.~4$A$).

As population declines, the number of groups exclusively formed by the residual plastic individuals in the $P$ state increase (shading in Figs.~4$A$-$B$). This situation drives the system to the ``recovery regime'' where inter-group variance makes the Simpson's paradox decisive once more (Fig.~4$D$). Note that this recovery dynamics, characteristic of the structure of commons, is enhanced by the heuristic at work: $P$ individuals that experienced groups of only $P$s keep contributing with high probability. The whole process stops the creation of $nP$s, expels cheaters and takes the population back to an all$P$ regime, an absorbing state of the system (Fig.~4.$B$). Once the population is uniquely constituted by $P$s, it remains in this homogeneous state until new cheaters arise.

If groups are relatively large, the mechanisms just described fail. While the reaction to cheater invasion is similar, the enrichment of all$P$ groups is delayed. Only when the population level becomes very low, these groups start to arise but, as we discussed earlier, this regime increases the chance of demographic extinctions. These collapses are not totally avoided in commons displaying higher $r$. In these cases, large group size $N$ and large efficiency $r$, we identified an alternative plastic heuristic that can assist sustainable growth.

\subsubsection*{Negative Plasticity is the Most Effective Strategy for Large Groups and High Efficiency} 
Individuals following a minimal-effort (plastic) heuristic are the ones that most strongly bring sustainability in commons structured in large groups, and where the supply of PG efficiently determines growth. These negative plastic producers present a $nP$ state unless the amount of PG in their latter interaction group is below a minimal threshold that could in the end impede growth (formally, they exhibit a large $q_>$, but small $q_<$ and $\theta$, see Fig.~1). Thus, a population constituted by negative plastic producers is constantly at low density, independently of the presence of cheaters.

The low-density regime is maintained as a dynamical equilibrium in which an excess of $P$s makes individuals to become $nP$, since many $P$s are observed in each group, while the successive lack of PG (and of $P$s in the groups) is compensated by showing again the $P$ state. In this scenario, the emergence of cheaters by mutation is indirectly controlled by the high abundance of $nP$ already in the resident population, which reduces in turn cheater presence and chance of invasion (Fig.~4$C$).

As negative plasticity strongly relies on the abundance (by default) of $nP$ agents, this heuristic requires that the PG produced by those that contribute must transform very efficiently into growth. For this reason, negative plasticity is successful only when $r$ is above a certain minimal value. Note also that, as compared to a population of positive plastic producers, the mechanisms of recovery do not rely so much on the temporal increase of inter-group variance (that brings the ``recovery regime'') but on the presence of a relatively constant and adequate inter-group diversity (Fig.~4$D$).

\section*{Discussion}
Communities whose growth depends on a PG contributed by their members present a fundamental instability associated to the emergence of  free-riders (cheaters) that do not contribute but use the accessible PG. This instability --at its core a problem of maintenance of cooperation-- produces direct ecological consequences, i.e., the collapse of the population. 

This ecological scenario immediately defines a characteristic ``environment" in which individuals following simple strategies are to ``solve" a precise task: attaining the sustainable growth of the collective. We analyzed this situation by considering limitations upon the decision-making capacities (Fig.~1) and also modifications of the specific attributes of the environment ($r$ and $N$) where decisions are taken~\cite{Simon1957}.

The analysis of the simplest heuristic,  constitutive production of PG, reveals the core ecological dynamics (Fig.~2$A$)~\cite{Hauert2006}. By avoiding production costs, cheaters can spread in a population of (constitutive) $P$s consequently reducing population density due to PG depletion. The resultant low densities induce the formation of between-group differences (groups dominantly constituted by $P$s or cheaters, Figs.~2$B$-$C$). This high inter-group variance causes individuals in groups dominantly composed by $P$s to receive larger payoffs, i.e., present higher replication rates. Differential growth leads to population recovery, as $P$ agents are the ones strongly contributing to the next generations (Simpson's paradox~\cite{Sober1984}, Fig.~2$A$-$B$). Low densities help then to promote $P$s (i.e., cooperation) in such structured populations.

Low densities originate a complementary ecological effect, when populations undergo demographic extinctions~\cite{Melbourne2008} instead of recovery. We captured these processes by quantifying the number of critical cascades --the number of times that the population is below a minimal density threshold. A decrease in $r$ or an increase of $N$ increases the number of critical cascades and the frequency of extinctions if PG is constitutively generated (Fig.~2$D$ and Fig.~$S2$). Hence, permanent production of PG does not always drive sustainable growth.

We introduced two additional heuristics. The simplest one, in which no external information is processed, consists in switching stochastically between contribution and non-contribution, i.e., individuals decide randomly to free-ride. This behaviour (defined by an opportune optimal switching rate, Fig.~3$A$-$B$) is generally more effective than constitutive production, but fails again when $r$ becomes smaller or $N$ larger (Figs.~$S4$ and $S5$).

The third heuristic is based on conditional contribution. We implement this by means of a simple sensing mechanism that allows individuals to estimate the composition of their recent interaction group and alter their behaviour accordingly. Modifying the two key structural attributes of the commons let us identify two contrasting conditional strategies.

For low $r$ (and sufficiently small group size $N$) positive plasticity is the most convenient strategy (Fig.~5$A$). This is related to its highly reactive response to cheaters. In fact, positive plastic producers stop producing PG when few cheaters (or few $nP$s) are detected in their earlier interaction group. This response immediately directs to minimal densities (Fig.~4$A$) and a successive strong recovery to the population carrying capacity (Fig.~4$B$). Interestingly, the reaction to cheaters invasions (consisting in the rapid increase of $nP$s) is promptly interrupted as result of the feedback between the threshold-like decision and group assortment of $P$s (created by the combination of low density and small group size, Fig.~4$A$). This example emphasizes how heuristics associated to limited information processing (the limitation corresponds in this case to the inability of individuals to distinguish the presence of cheaters from that of plastic producers in the $nP$ state) are still efficient due to the specific ecological structure where they are applied~\cite{Simon1957,Ostrom1994,Gigerenzer2002,Gigerenzer2011}.

Positive plasticity does not work when the commons is structured in large groups. In this case, negative plasticity emerges comparatively as a better strategy (Fig.~5$B$). This minimal effort behavior~\cite{Zipf1949} maintains the population in a dynamical equilibrium with the largest possible frequency of $nP$s that minimizes cheaters advantage but is compatible with population growth. Negative plasticity is in this sense an advanced version of stochastic production with the individual ability to switch back to a $P$ state when population density reaches critical values.

Thus, the use of different decision-making strategies clearly causes divergent sustainability outcomes when controlling for community structure (i.e, when both $N$ and $r$ is fixed, Fig.~5). One could further ask if these strategies are observed in natural scenarios (characterized by a PG dilemma).  We suggest that this is the case. Indeed, phenotypic noise, similar to the stochastic production strategy, is present in many bacterial communities, as a broad form of bet hedging~\cite{Beaumont2009} or in the stochastic expression of virulence factors, e.g.,~\cite{Ackermann2008}. Moreover, production of PGs such as extracellular polymeric substances --fundamental for biofilm communities-- is activated/terminated at high cell densities~\cite{Diggle2007,Nadell2008}; expression of bacteriocins is reduced when the population density is low by a quorum-sensing system~\cite{Ploeg2005}; generation of iron-scavenging pyoverdin molecules --iron being an essential public good in some environments-- is reduced when enough molecules are already present minimizing the ability of cheaters to invade~\cite{Kummerli2010}. In a different scale, that of human commons, various forms of conditional cooperation were also observed, with a variable degree of individual investment linked to group size~\cite{Rustagi2010,Fehr2003} (predicting minimal collaborative efforts in collective action as group size increases~\cite{Olson1965}, reminiscent of our results on negative plasticity; see also the ``hump shaped'' strategies in~\cite{Fischbacher2001}). 

That all the situations above correspond to very separate scenarios indicates that these heuristics could be fundamental ``building blocks''~\cite{Ostrom2005} in the assembly of this type of social arenas and, more broadly, in the maintenance of cooperation in structured populations (see Supplement for an extended discussion). Overall, these findings stress that, beyond the importance of structural factors, like PG efficiency and group structure, the sustainability of the commons should be understood as the appropriate integration of ecological dynamics and individual information-processing abilities.

\section*{Methods}
\subsection*{Public Good Games} 
Public good games are used to model social dilemmas in which the optimal behavior of the individual conflicts with the best outcome of the collective~\cite{Kagel1995}.  The simplest of these models is the one-shot public good game (also termed $N$-person prisoner's dilemma~\cite{Archetti2011}). This game is played by agents that can contribute (cooperators or producers, $P$) or not (defectors or non-producers, $nP$) to the PG in groups of size $N$. Contributing implies a cost $c$ to the agents. After the game, all contributions are summed, multiplied by an efficiency, or reward, factor $r$ (that determines the efficiency of the investments and the attractiveness of the PG) and redistributed to all individuals within the group, irrespectively of their contribution. Thus, if there are $i$ cooperators among the $N$ participants, the payoff for a defector is $icr/N$ while the payoff for a cooperator is $icr/N$ - $c$.

\subsection*{Computational Model} 
We used an agent-based computational model characterized by two distinct {\em stages} (Fig.~S1). In stage $I$, the population is structured in evenly-sized randomly formed groups in which the PG game is played. In stage $II$, after groups disappear, each individual chooses its successive phenotype (and then replicates) according to the group composition (and payoff) experienced in stage $I$. This model extends the original Hamilton's group selection model~\cite{Hamilton1975} to one in which individuals are able to dynamically change their phenotype according to previous experience.

Note that the PG game is specifically characterized by the parameters $N$, $r$, and $c$ (group size, efficiency and cost of the PG, respectively). We fix $c$=$1$ and $r$$<$$N$ (this second condition defines producers/cooperators as strong altruists~\cite{Fletcher2007}) and also the conditions of the initial population (constituted by a common pool of $k$ identical plastic agents in the $P$ state; $k$ is thus the maximal population size). The system is updated in a {\it sequential} way as follows (Supplement for further details): {\bf 1.} The common pool is divided in randomly formed groups of size $N$ (groups may contain individuals and empty spaces). {\bf 2.} In each one of the (non-empty) groups, a one-shot PG game is played. Agents in the $nP$ state and cheaters receive the payoff P$_{nP} = i\,c\,r/(i+j+w)$, while agents in the $P$ state receive the same payoff minus a cost, i.e., P$_P = $P$_{nP} -c $; with $i,j,w$ being the number of $P$s, $nP$s and cheaters in the group, respectively, and $i+j+w \leq N$. After the interaction the grouping of individuals is dissolved. {\bf 3.} Each plastic agent adjusts its state according to the relative abundance of $P$s that experienced in the group where it played the game and the triplet ($q_>, q_<, \theta$) as described in Fig.~1. {\bf 4.} Each individual replicates (creating an offspring) with a probability that is calculated  by dividing  its payoff by the maximal possible one (i.e., the payoff obtained by a $nP$, or equivalently a cheater, in a group of $N$-1 $P$s). Replication happens when the current total population presents less than $k$ individuals (i.e., there exits empty space).  For each replication, an individual generates an offspring that is a cheater with probability $\nu$ (i.e., a constitutive $nP$, see Fig.~1).  {\bf 5.} Individuals are removed with probability $\delta$ (death rate).  



\section*{Acknowledgments}
 This work was partially supported by CSIC ``programa junta para la ampliaci\'on de estudios'' JAEDOC15 (to M.C.) and by Spanish Ministry of Economy and Competitiveness BFU2001--24691 grant (to J.F.P.).

\section*{Contributions}
MC and JFP designed and performed research. Both authors wrote, reviewed and approved the manuscript.

\section*{Competing financial interests}
The authors declare no competing financial interests.


\begin{figure}
\centerline{\includegraphics[width=.9\textwidth]{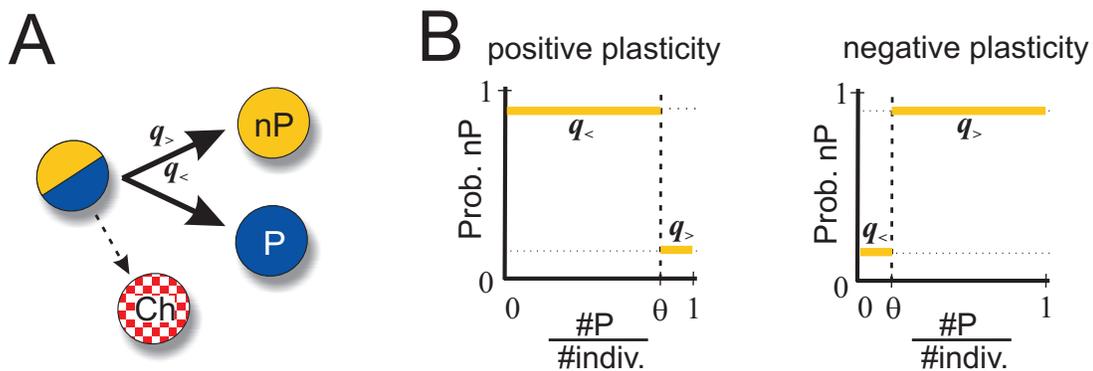}}
\caption{{\bf Individuals as plastic producers.} (A) A plastic producer can contribute (producer state, $P$) or not (non-producer state, $nP$) to the PG. This choice is made after each interaction and is conditioned to the specific group composition experienced, i.e., the ratio between the number of $P$s and the total number of individuals in the group. If this ratio is bigger, or equal, than a particular threshold value $\theta$, an agent becomes $nP$ with probability $q_>$ (if the ratio is smaller, it becomes $nP$ with $q_<$). Cheaters (Ch), i.e., agents that are permanently in the $nP$ state, arise from plastic producers by mutation. (B) Following this, a constitutive producer corresponds to the case $q_>=q_<=0$, while a stochastic producer exhibits nonzero $q_>= q_<$. {\em Positive} plastic producers (left panel) are characterized by a relatively large $\theta$. This implies that they hardly become $nP$ if the group they experienced was mostly constituted by $P$s (as they are also defined by a small $q_>$); if not, they express $nP$ with high probability (large $q_<$). In contrast, {\em negative} plastic producers (right panel) present a fairly small $\theta$. This means that they express $nP$ with high probability (as they are also defined by a large $q_>$) unless they were part of groups with few $P$s, in which case they hardly express $nP$ (small $q_<$). Note that the number of individuals ($\#$indiv.) can be less or equal than the group size ($N$).
}
\label{Fig1}
\end{figure}

\begin{figure*}
\centerline{\includegraphics[width=.95\textwidth]{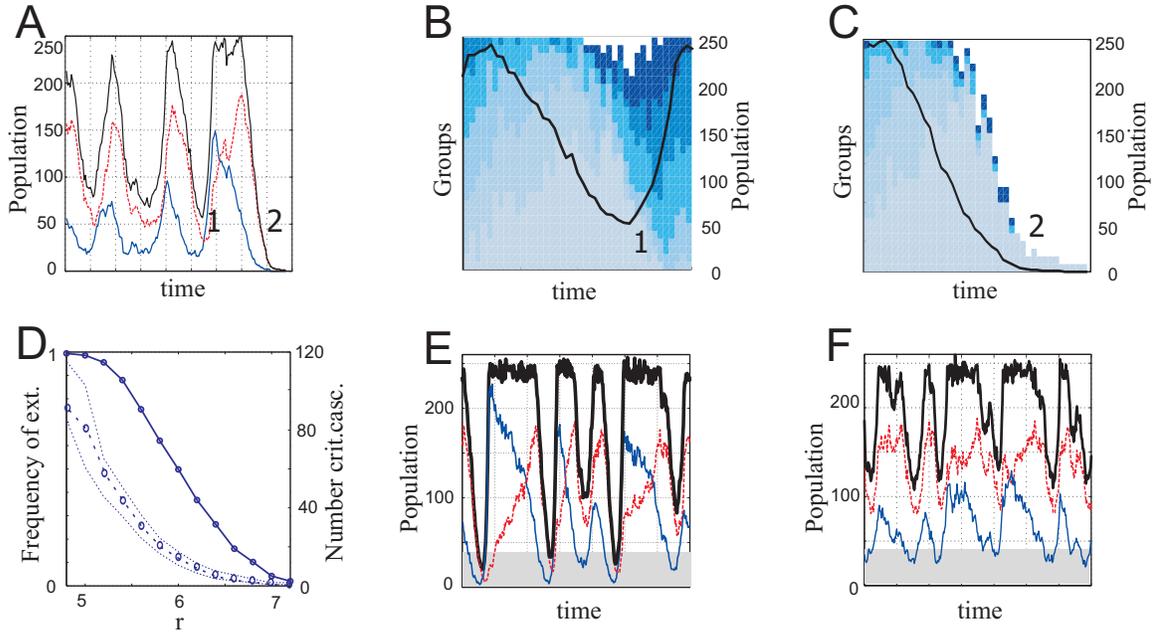}}
\caption{{\bf Growth based on constitutive producers is not sustainable for weakly efficient public good.} 
(A) Recovery and extinction in a population composed by constitutive $P$s and cheaters (color code as in Fig~1; total population in black). (B-C) Each group in the population (squares) is colored in different blue tones in a recovery (B) [around 1 in (A)] or an extinction (C) [around 2 in (A)] event. Group composition ranges from all $P$ (dark blue) to all cheaters (light blue); white denotes empty groups. Note the enrichment of groups with only $P$ immediately after each population decay. Unsuccessful replication of these initial groups causes population collapse. (D) Frequency of extinctions (continuous line) and median number of critical cascades (dashed line) as function of $r$. (E-F) Characteristic trace of regimes with low (E) and high (F) $r$. The population crosses more often the critical density region (highlighted in gray) at low $r$ for an equivalent time window.  A critical cascade is observed when the population crosses the critical density threshold, fixed to $30$ (i.e., = $k/10$).  Each point in (D) is the median (and 25/75$\%$ percentiles) of the average number of critical cascades obtained by considering all simulations that did not go extinct in $1000$ independent runs of $6 \times 10^5$ steps. Parameters: $k=300$, $N=10$, $\nu = 5 \times 10^{-6}$, $\delta=0.2$, $c=1$  (all panels); $r=4$ (A-B-C), $r=4.6$ (E) and $r=6.5$ (F).}
\label{Fig2}
\end{figure*}

\begin{figure}
\centerline{\includegraphics[width=.75\textwidth]{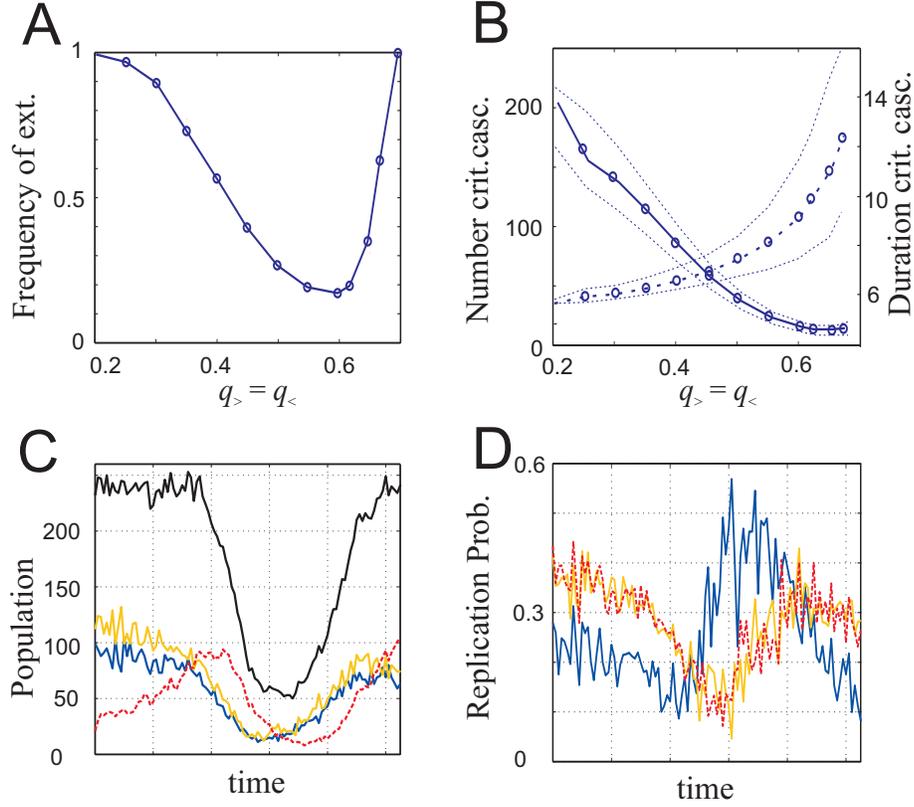}}
\caption{{\bf Stochastic producers can reduce extinctions.} (A) Frequency of extinctions as a function of the probability $q_>$ to stochastically express a $nP$ state ($q_> = q_<$, see Fig.~1). There exists an optimal $q_>$ that reduces extinctions by decreasing the number of critical cascades (continuous line in B) trading-off for their duration (dashed line in B). (C-D) Characteristic dynamics of a stochastic producer with optimal switching rate ($q_> =0.55$, color code as in Fig~1, black curve denotes total population). Note the limited replication rate of cheaters (generally below $\sim$$50\%$) even when population is high. Cheater advantage is stronger when agents constitutively produce PG (Fig.~S3).  A critical cascade is observed when the population crosses the critical density threshold, fixed to $30$ (i.e., = $k/10$), while the duration of a critical cascade is the average number of consecutive steps the population stays below the critical density threshold. Each point in (B) is the median (and 25/75$\%$ percentiles) obtained by considering all simulations that do not go extinct in $1000$ independent runs with $6\times 10^5$ steps. Parameters: $k=300$, $N=10$, $r=4$, $c=1$, $\delta=0.2$. 
} 
\label{Fig3}
\end{figure}

\begin{figure}
\centerline{\includegraphics[width=.75\textwidth]{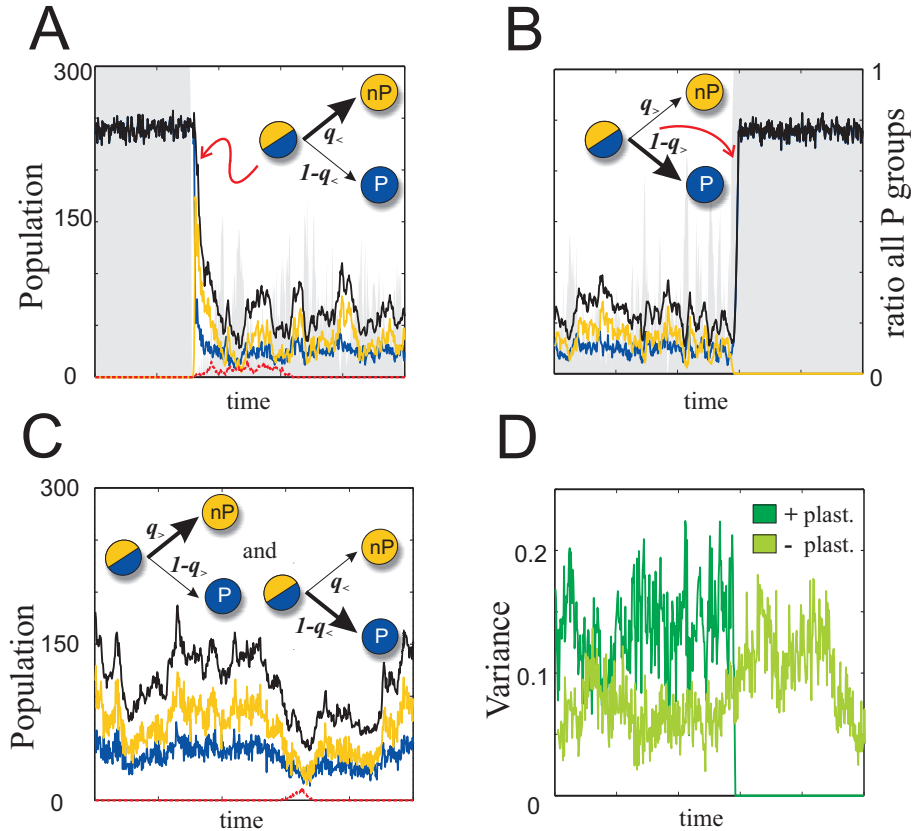}}
\caption{{\bf Positive and negative plastic producers include a sensing mechanism.} (A) Positive plastic producers expressing a $P$ state quickly switch to $nP$ as response to cheaters. This causes a fast reduction of $P$s overall (arrow thickness in inset cartoons denotes preferred individual decisions above/below $\theta$ with associated probabilities). After cheater invasion is stopped, the population exhibits coexistence of $P$ and $nP$ to finally evolve to an all$P$ scenario (B). Shading areas in (A-B) denote the relative amount of groups composed by only $P$s. (C) A population of negative plastic producers is characterized by its permanent low density favored by the constant presence of $nP$ which helps controlling cheaters invasion. (D) Positive plasticity transiently modifies the inter-group variance to control cheaters and stop the emergence of $nP$s. This contrasts with the relatively constant variance observed in a population of negative plastic producers (variances correspond to time series (B) and (C), respectively). Plastic producer definitions and color code as Fig.~1, with $q_>=0$, $q_<=0.7$ and $\theta =1$ for positive plasticity and $q_>=0.7$, $q_<=0$ and $\theta =0.1$ for negative; black curve describes total population. Parameters: $k=300$, $N=10$, $\delta=0.2$, $c=1$, $\nu=5 \times 10^{-6}$ (all panels); $r=2$ (A-B) and $r=2.2$ (C).}
\label{Fig4}
\end{figure}

\begin{figure}
\centerline{\includegraphics[width=.78\textwidth]{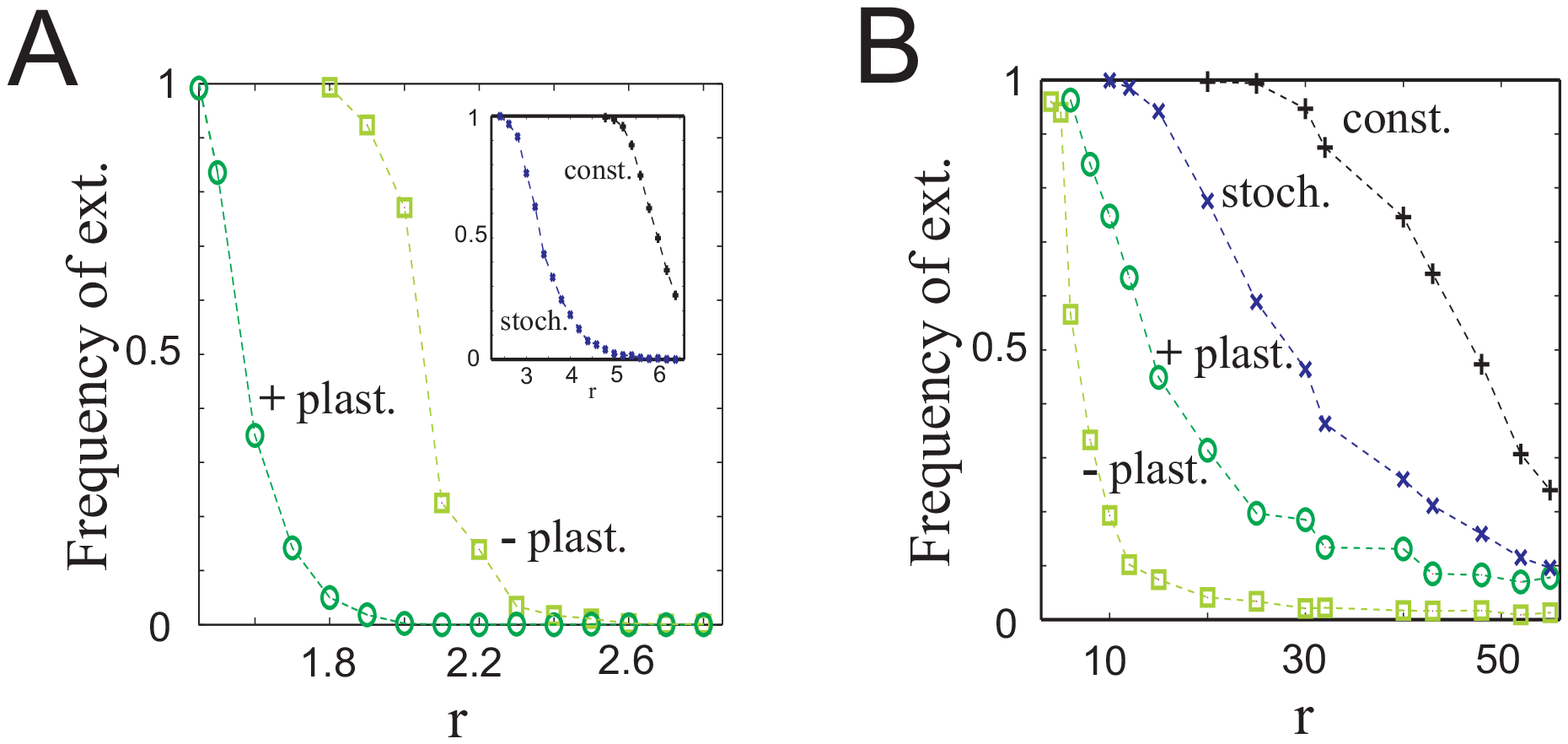}}
\caption{{\bf Individual strategies differentially manage sustainability in the commons.} Extinction frequencies observed by individuals exhibiting constitutive (black), stochastic (blue), and positive/negative (dark/light green) plasticity in a population structured in small [(A), $N$=10] or large [(B), $N=60$] groups (as a function of $r$). We considered $1000$ independent runs with $6 \times 10^5$ steps (Fig.~S8 and S9 considered a different $k$ and longer time series, respectively). For the curves corresponding to positive, negative plasticity and stochastic producers, the values plotted correspond to the minimal extinction frequency obtained by considering all possible instances of positive, negative and stochastic producers, respectively (see Figs.~S5--S7 and Supplement for details). Other parameters are $k=300$, $c=1$, $\delta=0.2$, $\nu=5 \times 10^{-6}$.}
\label{Fig5}
\end{figure}

\end{document}